\documentclass[11pt]{article}

\usepackage{amssymb}
\usepackage{amsmath}
\usepackage{mathrsfs}
\usepackage[dvipdfm]{hyperref}
\usepackage{color}

\begin{document}

\thispagestyle{empty}

\begin{flushright}
 KEK-TH-1250 
\end{flushright}

\vspace{10mm}
\begin{center}
{\Large \bf Janus field theories from multiple M2 branes}

\vspace{10mm}
{\large
Yoshinori Honma\footnote{E-mail address: 
yhonma@post.kek.jp},
Satoshi Iso\footnote{E-mail address: satoshi.iso@kek.jp},
Yoske Sumitomo\footnote{E-mail address: sumitomo@post.kek.jp}
and Sen Zhang\footnote{E-mail address: zhangsen@post.kek.jp}
}

\vspace{10mm}
 {\it Institute of Particle and Nuclear Studies, \\ High Energy Accelerator Research Organization(KEK) \\
 and \\ 
 Department of Particles and Nuclear Physics, \\
 The Graduate University for Advanced Studies (SOKENDAI), 
 \vspace{3mm} \\
Oho 1-1, Tsukuba, Ibaraki 305-0801, Japan} 

\end{center}

\vspace{20mm}
\ \ 

\begin{abstract}
Based on the recent proposal of  
${\cal N}=8$ superconformal  gauge theories
of  the multiple M2 branes, we derive (2+1)-dimensional 
supersymmetric Janus field theories with a
space-time dependent coupling constant.
 From the original Bagger-Lambert model, 
we get a supersymmetric field theory with 
a similar action to  the  $N$ D2 branes,
 but the coupling varies with the space-time 
 as a function of 
the light-cone coordinate, $g(t+x)$.
Half of the supersymmetries can be preserved.
We further investigate 
 the M2 brane action  deformed by 
 mass and Myers-like terms.
In this case, the final YM action is deformed by mass and
Myers terms and the coupling 
 behaves as $\exp(\mu x)$ where $\mu$ is a 
 constant mass parameter.
Weak coupling gauge theory is 
continuously changed to  strong coupling 
in the large $x$ region.

\end{abstract}

\newpage
\setcounter{page}{1}
\setcounter{footnote}{0}

\baselineskip 6mm
\section{Introduction}
\setcounter{equation}{0}
There has been a remarkable progress recently
in constructing ${\cal N}=8$ supersymmetric (2+1) field
theories with $SO(8)$ R-symmetry by
Bagger and Lambert \cite{Bagger:2007jr} and Gustavsson \cite{Gustavsson:2007vu}.
This model has 8 scalar fields and it is conjectured to be 
an effective field theory of multiple M2-branes in d=11.
An essential ingredient is the generalization
of the Lie algebraic structure  in ordinary
gauge theories to the Lie 3-algebras \cite{Bagger:2006sk}.
As expected  \cite{Schwarz:2004yj} the Lagrangian
contains a Chern-Simons term and a sextic
potential for scalars. 

The 3-algebraic structure is naturally 
expected for the M2 brane 
because the Schild form of the bosonic
membrane action is written in terms of 
the Nambu-Poisson bracket 
\begin{equation}
S \sim \int d^3 \sigma \  \{X^I, X^J, X^K \}^2, 
\end{equation}
where the Nambu-Poisson bracket \cite{Nambu:1973qe} is given by 
$ 
\{X^I, X^J, X^K \} = \epsilon_{ijk} \partial_i X^I
\partial_j X^J \partial_k X^K.
$ 
Then its quantum version must be written 
as 
\begin{equation}
S \sim {\rm Tr}  \  [X^I, X^J, X^K ]^2
\end{equation}
where the 3-algebra for the generators  $T^a$
is given by
$
[T^a, T^b, T^c ] = f^{abc}_{\ \ \ d} T^d.
$
The structure constant must obey the 
fundamental identity 
so that the action by Bagger-Lambert is invariant
under supersymmetry and gauge 
transformations.

Despite many efforts \cite{Awata:1999dz, Kawamura:2002yz,FigueroaO'Farrill:2002xg,Ho:2008bn}, 
the quantization of the
Nambu bracket is very hard and the only 
known example (satisfying the so called 
fundamental identity) was the algebra ${\cal A}_4$
\cite{Kawamura:2003cw} with 4 generators.
This is because
the requirement that the 3-algebra has a
positive definite metric is very strong.
It was conjectured \cite{Ho:2008bn}
and proved  \cite{Papadopoulos:2008sk,Gauntlett:2008uf} that
the only nontrivial positive definite 3-algebra is ${\cal A}_4$.
In order to circumvent this difficulty,
it was recently shown
\cite{Gomis:2008uv,Benvenuti:2008bt,Ho:2008ei}
that if we relax the condition of the positivity of the 
metric we can construct 3-algebras containing
the ordinary Lie algebra as a sub-algebra.
This is a remarkable progress.
The algebra contains 2 extra generators $T^{-1}$ and $T^0$
in addition to the generators of Lie algebra $T^i.$
(Here we use the convention of \cite{Ho:2008ei}.)
The 3-algebra for them is given by
\begin{eqnarray}
&&[T^{-1}, T^a, T^b] = 0,  \label{Nb1}\notag\\
&&[T^0, T^i, T^j] = f^{ij}_{\ \ k} T^k, \label{Nb2}\notag\\
&&[T^i, T^j, T^k] = f^{ijk} T^{-1}, \label{Nb3}
\end{eqnarray}
where $a,b = \{-1, 0, i \}$. $T^{i}$ are generators
of the ordinary Lie algebra with the structure
constant $f^{ij}_{\ \ k}$.
We can show that this satisfies the fundamental identity.
The metric $h^{ab}=Tr(T^a, T^b)$  is given by
\begin{eqnarray}
 &&\hbox{Tr}(T^{-1}, T^{-1}) = \hbox{Tr} (T^{-1}, T^{i}) = 0, \ \
  \hbox{Tr}(T^{-1},T^0) = -1, \nonumber \\
 &&\hbox{Tr}(T^0,T^i) = 0, \ \ \hbox{Tr}(T^0, T^0) = 0, \ \ 
 \hbox{Tr}(T^i,T^j) = h^{ij}.
\end{eqnarray}

Since the model contains negative metric, we may worry
that the model  based on the above 3-algebra
will contain ghost modes and they violate the unitarity 
of the theory.
The ghost modes are associated with the special 
components of the generators $T^{-1}$ and $T^0$.
Remarkably the authors of \cite{Gomis:2008uv,Benvenuti:2008bt,Ho:2008ei} showed
that the modes associated with the $T^{-1}$ generator
become Lagrange multipliers and the integration
gives a constraint
$
\partial^2 X^I_0=0
$
 for the other problematic modes
associated with $T^0.$
Then the would-be ghost modes can be decoupled
from the rest and the theory will be expected to become
unitarity.

The constraint $ \partial^2 X^I_0=0 $ is solved
as  $X^{I}_0= v \delta^I_{10} $ where
$v$ is a  constant\footnote{The idea of getting the D2-brane effective action by giving the
vev was originally given in \cite{Mukhi:2008ux}. }.
For a non-vanishing $v$, this breaks the $SO(8)$
R-symmetry to $SO(7)$.
After integrating 
non-dynamical modes of the gauge field, 
the  gauge theory action of $N$ D2 branes
is derived. The original model does not contain
any tunable parameter, but the value
of $v$ gives the coupling constant for the D2 brane
effective action.

In this paper we revisit the constraint equation.
The constraint equation $ \partial^2 X^I_0=0 $ is 
a massless wave equation and 
 a general function of the light cone coordinate,
$X^{I}_0=f(t+x) \ \delta^I_{10}$,  solves the constraint.
The integration of the non-dynamical gauge field
can be similarly performed and the resulting
theory becomes a (2+1)-dimensional {\it Janus gauge 
theory}. This breaks half of the original 16 supersymmetries.
In the Janus field theory, the coupling constant has the dependence on
coordinates.
Originally it was considered to be a dual of supergravity solutions with
a space-time dependent dilaton field \cite{Bak:2003jk}, and it has two
different ``faces'' at the boundary.
If there are two boundaries and there are different coupling constant
for each boundary, we should include interface terms which makes
 gauge couplings non-constant.
Supersymmetric field theories with the interface terms  are
constructed in \cite{D'Hoker:2006uv,Kim:2008dj,Gaiotto:2008sd}.

In order to fully quantize the model, we need to sum all the configurations
satisfying the constraint equations.  Towards the quantization and
proof of the unitarity, we consider general solutions
to the constraints, with no supersymmetries preserved, 
and see what kind of Janus field theory can be
derived around it. 

We will further investigate the mass deformation of the Bagger-Lambert action.
This model was studied by \cite{Gomis:2008cv,Hosomichi:2008qk} 
as a model of the matrix theory of type IIB plane waves.
 The deformed model has  desirable 
maximal supersymmetries as well as other bosonic symmetries.
In this case, the constraint equation is modified to 
$ (\partial^2-\mu^2 )X^I_0=0 $ and the solution of this 
constraint is given by 
$X^I_0=\exp(\mu x) \delta^I_{10}$ where $x$ is a space
direction. This preserves half of the original
supersymmetries.
The non-dynamical gauge modes can be integrated out again
and the theory becomes a supersymmetric Janus field
theory with a  Myers-term added.
The gauge coupling constant changes from weak to strong
as we move along the coordinate $x$ from $-\infty$ to $+\infty.$

The organization of the paper is as follows.
In section 2, we first review the Bagger-Lambert model
based on the realization of 3-algebra with a negative component of the metric. 
We also comment that the constraint equation has more generic
solutions with the coupling constant varying with the space-time coordinates
as a function of the light-cone coordinate.
In section 3, we  extend the model including a mass and Myers-like term
and investigate the model similarly.

There are many other interesting developments of  multiple M2-branes \cite{others}.

\section{Bagger-Lambert model}
\setcounter{equation}{0}
\subsection{Brief review of BL model}
We first briefly review the Bagger-Lambert action
and its symmetry properties.
It is a (2+1)-dimensional nonabelian gauge theory with ${\cal N}=8$ 
supersymmetries. It contains 8 real scalar fields 
$X^I=\sum_a X^I_a T^a, \ I=3,...,10$,
gauge fields $A^\mu=\sum_{ab}A^\mu_{ab} T^a 
\otimes T^b, \mu=0,1,2$ with two internal indices
and 11-dimensional Majorana spinor fields 
$\Psi= \sum_a\Psi_a T^a$ with a chirality
condition $\Gamma_{012}\Psi=\Psi.$
The action proposed by Bagger and Lambert 
is given by
\begin{eqnarray}
{\cal L} = -\frac{1}{2} \hbox{Tr}(D^{\mu}X^I, D_{\mu} X^I) 
           + \frac{i}{2} \hbox{Tr}(\bar\Psi, \Gamma^{\mu}D_{\mu}\Psi) 
           +\frac{i}{4} \hbox{Tr}(\bar\Psi, \Gamma_{IJ} [X^I, X^J, \Psi])
           -V(X) + {\cal L}_{CS}. \label{BLaction}
\end{eqnarray}
where
$D_{\mu}$ is the covariant derivative defined by:
\begin{eqnarray}
 (D_{\mu} X^I)_a = \partial_{\mu} X^I_a - f^{cdb}_{\ \ \ a} A_{\mu cd}(x) X^I_b.
\end{eqnarray}
$V(X)$ is a  sextic potential term 
\begin{eqnarray}
 V(X) = \frac{1}{12} \hbox{Tr}([X^I,X^J,X^K],[X^I,X^J,X^K]),
\end{eqnarray}
and the Chern-Simons term for the gauge potential is given by
\begin{eqnarray}
 {\cal L}_{CS} = \frac{1}{2} \epsilon^{\mu\nu\lambda} (f^{abcd} A_{\mu
  ab} \partial_{\nu} A_{\lambda cd} + \frac{2}{3}f^{cda}_{\ \ \ g} f^{efgb}
  A_{\mu ab} A_{\mu cd} A_{\lambda ef}).
\end{eqnarray}
This action is invariant under the SUSY transformation
\begin{eqnarray}
 \delta X^I_a &=& i \bar{\epsilon} \Gamma^I \Psi_a, \nonumber \\
 \delta \Psi_a &=& D_{\mu} X^I_a \Gamma^{\mu} \Gamma^{I} \epsilon -
  \frac{1}{6} X^I_b X^J_c X^K_d f^{bcd}{}_a \Gamma^{IJK}\epsilon,
  \nonumber \\
 \delta \tilde{A}_{\mu \ a}^{\ b} &=& i\bar{\epsilon} \Gamma_{\mu} \Gamma_{I}
  X^I_c \Psi_d f^{cdb}{}_a, \hspace{2em} \tilde{A}_{\mu \ a}^{\ b} \equiv
  A_{\mu cd} f^{cdb}{}_a,
\end{eqnarray}
and the gauge transformation
\begin{eqnarray}
 \delta X^{I} &=& \Lambda_{ab} [T^a,T^b,X^I], \nonumber \\
 \delta \Psi &=& \Lambda_{ab} [T^a,T^b,\Psi], \nonumber \\
 \delta \tilde{A}^{\ b}_{\mu \ a} &=& D_{\mu} \tilde{\Lambda}^{b}_{\ a},
 \hspace{2em} \tilde{\Lambda}_{\ a}^{\ b} \equiv
  \Lambda_{cd} f^{cdb}{}_a,
\end{eqnarray}
provided that the triple product  $[A,B,C]$ has the fundamental 
identity and $\hbox{Tr}$ satisfies the  property discussed in the next 
subsection. 
The most peculiar property of the model is that the gauge
transformation and the associated gauge fields
have two internal indices. This must come from the 
volume preserving diffeomorphism of the membrane
action \cite{de Wit:1988ig,Bergshoeff:1987cm} but the concrete
realization of the gauge symmetry from the supermembrane
action is not yet clear. 
\subsection{A specific realization of 3-algebra}
This theory is based on an antisymmetric 3-algebraic structure ${\cal G}$
with generators $T^a$
\begin{eqnarray}
 [T^a,T^b,T^c] = f^{abc}_{\ \ \ d} T^d.
\end{eqnarray}
Here we take the specific realization of the 3-algebra
containing the ordinary Lie algebra as a sub-algebra.
The most fundamental identity of the algebra is
the generalized Jacobi identity.
It is called the ``fundamental identity'' and given by
\begin{eqnarray}
 [T^a,T^b,[T^c,T^d,T^e]] = [[T^a,T^b,T^c],T^d,T^e] +
  [T^c,[T^a,T^b,T^d],T^e] + [T^c,T^d,[T^a,T^b,T^e]].
\end{eqnarray}
If this identity holds, we can show that the gauge 
transformations generated by $T^a \otimes T^b$
 form Lie algebra\footnote
 {Strictly speaking, $\tilde{T}^{ab}$ satisfies ordinary Lie algebras
only when they act on $X$. If we write the commutation relations of $\tilde{T}^{ab}$
without acting on $X$, they are not necessarily associative and contain
3-cocycles.}.
Namely, if we write $\tilde{T}^{ab} X = [T^a, T^b, X]$,
a commutator closes among the generators $\tilde{T}^{ab}$;
\begin{eqnarray}
 [\tilde{T}^{ab}, \tilde{T}^{cd}]X &=& [T^a,T^b, [T^c,T^d,X]] - [T^c,T^d, [T^a,T^b,X]] 
\nonumber  \\ &=& 
[[T^a,T^b,T^c],T^d,X]+ [T^c,[T^a,T^b,T^d], X]  
\nonumber \\
&=& (f^{abc}_{ \ \ \ e} \tilde{T}^{ed} + f^{abd}_{\ \ \ e} \tilde{T}^{ce}) X.
\end{eqnarray}

A specific choice of the 3-algebra satisfying the fundamental identity
is given by \cite{Gomis:2008uv,Benvenuti:2008bt,Ho:2008ei}. 
It contains an ordinary set of Lie algebra generators
as well as two extra generators
$T^{-1}$ and $T^{0}$. The algebra is given by
\begin{eqnarray}
&&[T^{-1}, T^a, T^b] = 0,  \label{Nb1}\notag\\
&&[T^0, T^i, T^j] = f^{ij}_{\ \ k} T^k, \label{Nb2}\notag\\
&&[T^i, T^j, T^k] = f^{ijk} T^{-1}, \label{Nb3}
\end{eqnarray}
where $a,b = \{-1, 0, i \}$. $T^{i}$ is a generator
of the  Lie algebra and $f^{ij}_{\ \ k}$ is its structure constants.
Here $T^{-1}$ is the central generator meaning  that its triple product
with any other generators vanishes.
$T^{0}$ is also special since it is not generated by the
3-algebra and does not appear in the right hand side of the triple product.
One can easily check that this triple product satisfies the fundamental
identity. 
In order to construct a gauge invariant field theory Lagrangian,
we need the trace operation with the identity
\begin{eqnarray}
 \hbox{Tr}([T^a,T^b,T^c],T^d) + \hbox{Tr}(T^c,[T^a,T^b,T^d]) = 0.
\end{eqnarray}
After a suitable redefinition of generators, 
such a trace can be given by
\begin{eqnarray}
 &&\hbox{Tr}(T^{-1}, T^{-1}) = \hbox{Tr} (T^{-1}, T^{i}) = 0, \ \
  \hbox{Tr}(T^{-1},T^0) = -1, \nonumber \\
 &&\hbox{Tr}(T^0,T^i) = 0, \ \ \hbox{Tr}(T^0, T^0) = 0, \ \ 
 \hbox{Tr}(T^i,T^j) = h^{ij}.
\end{eqnarray}
If we define $f^{abcd}$ as
$f^{abcd} = f^{abc}{}_e h^{ed}$,
 $f^{abcd}$ is totally antisymmetry. 
 
The above construction of the 3-algebra contains the 
ordinary Lie algebra as a sub-algebra.
The generators of the gauge transformation
can be classified into 3 classes.
\begin{itemize}
\item{${\cal I}$}=$\{ T^{-1} \otimes T^a , a=0,i\}$
\item{${\cal A}$}=$\{T^0 \otimes T^i  \}$
\item{${\cal B}$}=$\{ T^i \otimes T^j \}$
\end{itemize}
Then it is easy to show that 
\begin{equation}
[{\cal I}, {\cal I}]=[{\cal I}, {\cal A}]=[{\cal I} ,{\cal B}]=0, 
\ [{\cal A}, {\cal A}]={\cal A}, \ [{\cal A}, {\cal B}] = {\cal B},
\ [{\cal B},{\cal B}] = {\cal I}
\end{equation}
and hence the generators of ${\cal A}$ form a sub-algebra,
which can be identified as the Lie algebra of $N$ D2-branes. 
\subsection{BL model  to D2 branes} 
In the specific realization of the 3-algebra, 
we can decompose the modes of the fields as
\begin{eqnarray}
 X^I &=& X^I_0 T^0 + X^I_{-1} T^{-1} + X^I_i T^i, \notag\\
\Psi &=& \Psi_0 T^0 + \Psi_{-1} T^{-1} + \Psi_i T^i, \notag\\
A_{\mu} &=& T^{-1} \otimes A_{\mu(-1)} - A_{\mu(-1)} \otimes T^{-1}
 \nonumber \\ & & + A_{\mu 0j} T^0 \otimes T^j - A_{\mu j0}
 T^j \otimes T^0 + A_{\mu ij} T^i\otimes T^j.  
\end{eqnarray} 
It will be convenient to define the following fields as in \cite{Ho:2008ei}
\begin{eqnarray}
 &&\hat{X^I} = X^I_{i} T^i, \hspace{2em}
  \hat{\Psi} = \Psi_{i} T^i \nonumber \\
 &&\hat{A}_{\mu} = 2 A_{\mu 0 i} T^i, \hspace{1em} B_{\mu} =
  f^{ij}{}_k A_{\mu ij} T^k.
\end{eqnarray}
The gauge field $A_{\mu (-1)}$ is decoupled from
the action and we drop it in the following discussions.
The gauge field $\hat{A}_{\mu}$ is associated with the 
gauge transformation of the sub-algebra ${\cal A}$.
Another gauge field $B_\mu$ will play a role of the 
$B$-field of the BF theory and can be integrated out.
With these expression the Bagger-Lambert action (\ref{BLaction}) can be
rewritten as
\begin{eqnarray}
 {\cal L} &=& \hbox{Tr}
   \left( - \frac{1}{2}(\hat{D}_{\mu} \hat{X}^I - B_{\mu} X_0^I)^2 +
   \frac{i}{2} \bar{\hat{\Psi}} \Gamma^{\mu} \hat{D}_{\mu} \hat{\Psi} +
   i \bar{\Psi}_0 \Gamma^{\mu} B_{\mu} \hat{\Psi} +
   \frac{1}{4}(X_0^K)^2 ([\hat{X}^I,\hat{X}^J])^2 \right. \nonumber \\
  && \left. - \frac{1}{2} (X_0^I [\hat{X}^I,\hat{X}^J])^2
      -\frac{1}{2}\bar{\Psi}_0 \hat{X}^I
      [\hat{X}^J,\Gamma_{IJ}\hat{\Psi}]
      + \frac{1}{2}\bar{\hat{\Psi}}X^I_0[\hat{X}^J,\Gamma_{IJ}\hat{\Psi}]
      + \frac{1}{2} \epsilon^{\mu\nu\lambda} \hat{F}_{\mu\nu}
      B_{\lambda} \right.
    \nonumber \\
  & & \left.
- \partial_{\mu} X^I_0\  B_{ \mu} \hat{X}^I
 \right) 
+ {\cal L}_{gh},
\end{eqnarray}
where the ghost term is 
\begin{eqnarray}
 {\cal L}_{gh} =   (\partial_{\mu} X^I_0)(\partial^{\mu} X^I_{-1}) - i \bar{\Psi}_{-1}
  \Gamma^{\mu} \partial_{\mu} \Psi_0.
\end{eqnarray}
The covariant derivative and the field strength
\begin{eqnarray}
 \hat{D}_{\mu} \equiv \partial_{\mu} \hat{X}^I + i
  [\hat{A}_{\mu},\hat{X}^I], \hspace{1em}
  \hat{D}_{\mu} \Psi \equiv \partial_{\mu} \hat{\Psi} + i
  [\hat{A}_{\mu},\hat{\Psi}], \hspace{1em} \hat{F}_{\mu\nu} =
  \partial_{\mu}\hat{A}_{\nu} - \partial_{\nu} \hat{A}_{\mu} + i
  [\hat{A}_{\mu}, \hat{A}_{\nu}]
\end{eqnarray}
are the ordinary covariant derivative and field strength for the 
sub-algebra ${\cal A}.$
As emphasized in \cite{Gomis:2008uv,Benvenuti:2008bt,Ho:2008ei}, a coupling constant
can be always absorbed by the field redefinition and there 
is no tunable parameters in this model.

The supersymmetry transformations for each mode are
given by 
\begin{eqnarray}
\delta X^I_0 &=& i\bar{\epsilon} \Gamma^I \Psi_0, \notag\\
\delta X^I_{-1} &=& i \bar{\epsilon} \Gamma^I \Psi_{-1}, \notag\\
 \delta \hat{X}^I &=& i\bar{\epsilon} \Gamma^I \hat{\Psi}, \notag\\
 \delta \Psi_0 &=& \partial_{\mu} X^I_0 \Gamma^{\mu} \Gamma^I
  \epsilon, \notag\\
 \delta \Psi_{-1} &=& \{\partial_{\mu} X^I_{-1} - \hbox{Tr}(B_{\mu},
  \hat{X}^I)\} \Gamma^{\mu} \Gamma^I \epsilon + \frac{i}{6}
  \hbox{Tr}(\hat{X}^I, [\hat{X}^J,\hat{X}^K])\Gamma^{IJK} \epsilon, \notag\\
 \delta \hat{\Psi} &=& \hat{D}_{\mu} \hat{X}^I \Gamma^{\mu} \Gamma^I
  \epsilon -B_{\mu}X^I_0 \Gamma^{\mu} \Gamma^I \epsilon + 
  \frac{i}{2}X^I_0[\hat{X}^J,\hat{X}^K] \Gamma^{IJK}\epsilon, \notag\\
 \delta \hat{A}_{\mu} &=& i\bar{\epsilon} \Gamma_{\mu} \Gamma_I (X^I_0
  \hat{\Psi} - \hat{X}^I\Psi_0), \notag\\
 \delta B_{\mu} &=& \bar{\epsilon} \Gamma_{\mu} \Gamma_{I}
  [\hat{X}^I,\hat{\Psi}].
\end{eqnarray}

Here note that 
$X^I_{-1}$ and $\Psi_{-1}$ appear only linearly in the  Lagrangian
and thus they are  Lagrange multipliers. 
By integrating out these fields, we have the following 
constraints for the other problematic fields associated with
$T^0$;
\begin{eqnarray}
 \partial^2 X^I_0 = 0, \hspace{2em} \Gamma^{\mu} \partial_{\mu} \Psi_0 = 0.
\end{eqnarray}
This should be understood as a physical state condition 
$\partial^2 X^I_0|phys \rangle=0$.
In the path integral formulation, these constraints appear
as a delta function  $\delta(\partial^2 X^I_0)$
and those fields are constrained to satisfy 
the massless wave equations.
In order to fully quantize the theory, we need to sum
all the solutions satisfying the constraints, but we here
take a special solution to the constraint equations 
and see what kind of field theory can be  obtained.

The simplest solution is given by 
\begin{eqnarray}
 X^I_0 = v \ \delta^I_{10}, \hspace{2em} \Psi_0 = 0,
\end{eqnarray}
where $v$ is some constant.  
This solution was considered in \cite{Gomis:2008uv,Benvenuti:2008bt,Ho:2008ei} and 
preserves all the 16 supersymmetries, 
the gauge symmetry generated by the subalgebra ${\cal A}$,
and $SO(7)$ R-symmetry rotating $X^A, \ A=3,...,9.$ 
Another interesting solution is given by
\begin{eqnarray}
 X^I_0 = v(x^0+x^1) \delta^I_{10}\ , \hspace{2em} \Psi_0 = 0
\end{eqnarray}
where $v(x^0+x^1)$ is an arbitrary function on the light cone
coordinate. 
As we see the supersymmetry transformation for $\Psi_0$,
\begin{equation}
\delta \Psi_0 = \partial_\mu X^I_0 \Gamma^\mu \Gamma^I \epsilon,
\end{equation}
the solution $X^I_0 = v(x^0+x^1) \delta^I_{10}$ 
preserves half of the supersymmetries.

In both cases, if we fix 
the fields $X^I_0$ and $\Psi_0$ as above, 
we can integrate over the  gauge field
$B_\mu$ and obtain the effective action for $N$ D2 branes\footnote
{The fermion here is a 32 component spinor satisfying 
$\Gamma_{012}\Psi=\Psi$. In order to  recover the ordinary notation 
for D2 branes, we  rearrange it
 as $\tilde{\Psi}=(1+\Gamma_{10}) \Psi$.
Then it satisfies $\Gamma_{10} \tilde{\Psi}=\tilde{\Psi}$ and the action
is written in the usual form (no $\Gamma_{10}$ in the last term). }
\begin{eqnarray}
 {\cal L} = {\rm Tr}\left[-\frac{1}{2} (\hat{D}_{\mu} \hat{X}^A)^2 + \frac{1}{4} v^2
  [\hat{X}^A,\hat{X}^B]^2 + \frac{i}{2}
  \bar{\hat{\Psi}} \Gamma^{\mu}
  \hat{D}_{\mu} \hat{\Psi}
  - \frac{1}{4v^2} \hat{F}_{\mu\nu}^2 +
  \frac{1}{2} v \bar{\hat{\Psi}}  [\hat{X}^A, \Gamma_{10,A} \hat{\Psi}]
                    \right],
\end{eqnarray}
where $A,B=3,\cdots,9$. 
The coupling $v$ is given by the vev of 
$X^{10}_0$ and it is either
a constant or an arbitrary function on the light-cone $v(x^0+x^1).$
This may be identified as the compactification radius of 11-th direction in M-theory;
$v = 2\pi g_s l_s.$
The supersymmetric  YM theories with a space-time
dependent  coupling are known
as Janus field theories
and originally considered to 
be a dual of supergravity solutions with 
space-time dependent dilaton fields \cite{Bak:2003jk}.

A salient feature is that the 10-th spacial
fields $X^{10}$ completely disappear from the Lagrangian
by integrating out the redundant gauge field $B_\mu$. 
It is interesting that Janus field theories are naturally 
obtained from the Bagger-Lambert field theories.

The $v \rightarrow 0$ limit cannot be taken after integrating
the redundant gauge field $B_\mu.$
In the case of vanishing $v$, the Lagrangian is simply given by
\begin{eqnarray}
 {\cal L} = {\rm Tr}\left[-\frac{1}{2} (\hat{D}_{\mu} \hat{X}^I)^2 
 + \frac{i}{2} \bar{\hat{\Psi}} \Gamma^{\mu}
  \hat{D}_{\mu} \hat{\Psi} \right]
\end{eqnarray}
with a constraint $\hat{F}_{\mu \nu}=0.$
The action is of course invariant under the full $SO(8)$ R-symmetry.

\subsection{Janus field theory with Dynamical coupling}
In the previous subsection, we have fixed the solution of the
constraint equations. But in the quantization of the Bagger-Lambert 
model, the solutions should be summed in the path integral.
So we will consider more general solutions in this subsection.
After integrating the modes associated with the $T^{-1}$ generator,
the partition function becomes
\begin{equation}
Z = \int {\cal D}X_0^I {\cal D}\Psi_o  {\cal D}B_\mu  {\cal D}\hat{X}^I 
 {\cal D}\hat{\Psi}   {\cal D}A_\mu \ 
\delta(\partial^2 X_0^I) \ 
\delta(\Gamma^{\mu} \partial_\mu \Psi_0) \ 
e^{i S(X_o^I, \Psi_0, B_\mu, \hat{X}^I, \hat{\Psi}, A_\mu)}.
\end{equation}
The integrations over $X^I_0$ and $\Psi_0$ are constrained to 
obey the massless wave equations and can be expanded as
\begin{equation}
X^{I}_0 = \sum_n c_n^I f_n(x), \ \ \  \Psi_0=\sum_n  b_n u_n(x)
\end{equation}
where $f_n(x), u_n(x)$ are complete sets of functions satisfying the massless 
wave equations. Then the integration over $X^I_0$ and $\Psi_0$ can be
reduced to integrations over $c_n^I$ and $b_n.$

Let us now choose a general  solution $(X_0^I=v^I(x), \Psi_0)$ to the constraints
 and expand the action around it. 
In this case all the supersymmetries are generally broken if we fix $v^I$ and $\Psi_0$.
Inserting this general solution into the action, 
terms including the $B_\mu$ gauge field are given by
\begin{equation}
-\frac{1}{2} (\hat{D}_{\mu} \hat{X}^I - B_{\mu} X_0^I)^2 + 
  i \bar{\Psi}_0 \Gamma^{\mu} B_{\mu} \hat{\Psi} +
  \frac{1}{2} \epsilon^{\mu\nu\lambda} \hat{F}_{\mu\nu}
      B_{\lambda} 
     - \partial_{\mu} X^I_0  B_{ \mu} \hat{X}^I  .
\end{equation}
The integration over the $B_\mu$ gauge field can be similarly performed.
It is convenient to  introduce the locally defined projection operator
\begin{equation}
P_{IJ}(x) = \delta_{IJ} - \frac{v_I v_J}{v^2},
\end{equation}
This operator satisfies $P^2=P$ and $P_{IJ}v^J =0$.
In the simplest case considered in the previous subsection, $v^I = v(t+x) \delta^I_{10}$,
this projects out the 10-th direction if it acts on $\hat{X}^I$.
Generally, the direction removed is dependent on the space-time position.
 
After integrating over the $B_\mu$ field, the Lagrangian becomes
${\cal L}_{Janus} ={\cal L}_{0} + {\cal L}'$ where
\begin{eqnarray}
{\cal L}_{0} &=&
{\rm Tr}\left[  -\frac{1}{2} (\hat{D}_{\mu} Y^I)^2 + \frac{1}{4} v^2
  [Y^I, Y^J]^2 + \frac{i}{2}
  \bar{\hat{\Psi}} \Gamma^{\mu}
  \hat{D}_{\mu} \hat{\Psi}  +
  \frac{1}{2}  \bar{\hat{\Psi}}  [Y^I, (v^J \Gamma_{J}) \Gamma_{I} \hat{\Psi}]
  \right.  \nonumber \\ 
&& \left.
+\frac{1}{2 (v^I)^2} \big(  
\frac{1}{2} \epsilon^{\mu \nu \lambda} \hat{F}_{\nu \lambda} + i \bar{\Psi}_0 \Gamma^\mu \hat{\Psi}
- 2  Y_I \partial^\mu v^I  \big) ^2 
-\frac{1}{2} \bar{\Psi}_0 \Gamma_{IJ} \hat{\Psi} [Y^I,Y^J]
                    \right],
\label{JanusFT0}
\\
{\cal L}' &=& \frac{1}{v^2}  {\rm Tr} \left[
\left( \bar{\Psi}_0 \Gamma_{I}
(v^J \Gamma_{J}) [Y^I, \hat{\Psi}] - i \bar{\Psi}_0 \Gamma_{\mu} \hat{D}_\mu \hat{\Psi}
\right)
(v^K \hat{X}^K) \right].
\label{JanusFT}
\end{eqnarray}
Here $I, J=3, \cdots, 10$ and  
we have defined a new scalar field $Y^I = P_{IJ} \hat{X}^J$
with  7 degrees of freedom. In spite of it, the action has $SO(8)$ invariance if
$v^I$ and $\Psi_0$ also transform under it. Also note that $Y^I$ is invariant
under the gauge transformations associated with $B_\mu$ gauge fields.
Is is also interesting to notice that the action will have a generalized conformal 
symmetry \cite{Jevicki} even with the dimensionful coupling 
because it is  a dynamical variable here. 
This  may have its origin 
in the conformal symmetry of M2 branes. 
In this sense, the reduced action is not exactly the same as the ordinary D2 brane
effective action with a fixed gauge coupling.
This issue is now under investigations.

This is a Janus field theory whose coupling varies with space-time. 
The Lagrangian ${\cal L}_{YM}$ contains only the projected scalar field $Y^I$.
On the other hand, in the presence of $\Psi_0$, the scalar field
$(v^I \hat{X}^I)$ does not decouple from the Lagrangian
${\cal L}'$.
If we can set $\Psi_0=0$, ${\cal L}'$ vanishes
and the resultant Lagrangian is given by 
a similar form to the ordinary
Super Yang-Mills Lagrangian, but the  kinetic term of the 
gauge field $\hat{F}_{\mu \nu}$ is modified 
to $\hat{F}_{\mu\nu} + 2 \epsilon_{\mu\nu\rho} Y_I \partial^\rho v^I.$
All the supersymmetries are generally broken if we fix 
one solution to the constraint equations of $(X^I_0(x), \Psi_0)$
as above. 

By using the above calculation, the partition function can be simply
rewritten as
\begin{equation}
Z = \int \prod_n dc_n^I \ db_n \  W(v^I)
\int    {\cal D}\hat{X}^I 
 {\cal D}\hat{\Psi}   {\cal D}A_\mu \ 
e^{i S_{Janus}(\hat{X}^I, \hat{\Psi}, A_\mu; v^I(x), \Psi_0)}.
\end{equation}
Here $W(v^I) \sim ((v^I)^2)^{-3/2}$ came
from the integration over the $B_\mu$ field.
It is a sum of Janus field theories. The coupling constant $v^I$
is dynamical and varies with space-time coordinates.
It is constrained to satisfy the massless equations. 
If we fix the ``slow'' variable $v$ and  perform the path integration over the 
other ``fast'' variables first, then we can 
get an effective action for the dynamical coupling  $v^I.$ 
This will determine the most stable configuration of $v^I(x)$,
and accordingly one of the Janus gauge theory
with the most stable coupling is determined. 
If the variable $v^I$ fluctuates rapidly and cannot be
considered as a slow variable, 
the theory becomes very different from the ordinary gauge theory
with a fixed (either constant or varying)  gauge coupling. 
This may be related to the dynamical determination of 
the compactification radius of 11-th direction in M-theory.

Finally we would like to comment on the unitarity of the Bagger-Lambert theory.
If we fix one solution to the constraints, each theory behaves regularly if
the coupling constant does not vary  drastically.
The quantization of the coupling is very difficult, but since it is
not a propagating mode, it will not violate the unitarity of the
theory.
However the unitarity should be more carefully analyzed.
\section{Mass deformation and  Janus solutions}
\setcounter{equation}{0}
\subsection{Mass deformation of BL}
The BL model in the previous section gives 
a familiar effective action of $N$ D2 branes
with either a  constant or a varying coupling. 
(For general solutions, the kinetic term of the gauge field
contains a non-familiar term of $Y_I \partial^\mu v^I$.)

In this section we start from a 
 mass deformed Bagger-Lambert action given by
\cite{Gomis:2008cv,Hosomichi:2008qk} and show that  supersymmetric 
Janus field theories with  a Myers-term are obtained.

One parameter deformation of the Bagger-Lambert action
preserving the full supersymmetries is given by adding 
the following mass and  flux terms to the original Lagrangian.
The mass term is given by
\begin{eqnarray}
 {\cal L}_{mass} = -\frac{1}{2} \mu^2 \hbox{Tr}(X^I,X^I) + \frac{i}{2}
  \mu \hbox{Tr} (\bar{\Psi}\Gamma_{3456},\Psi),
\end{eqnarray}
and a flux term is 
\begin{eqnarray}
 {\cal L}_{flux} = -\frac{1}{6} \mu \epsilon_{EFGH}
  \hbox{Tr}([X^E,X^F,X^G],X^H) -\frac{1}{6}\mu
  \epsilon_{E^{'}F^{'}G^{'}H^{'}}
  \hbox{Tr}([X^{E^{'}},X^{F^{'}},X^{G^{'}}],X^{H^{'}}).
\end{eqnarray}
Here $E,F,G,H = 3,4,5,6$ and $E^{'},F^{'},G^{'},H^{'} = 7,8,9,10$. This
action is invariant under the original gauge transformation and the
deformed SUSY transformation 
\footnote{To give a rigorous proof of the closure of the supersymmetry, 
we should check the Jacobi identity of
$[Q,\{Q,Q\}]$ (appendix E of \cite{Lin:2005nh})
because there are non-central terms, i.e.
$SO(4) \times SO(4)$ rotation term, in the
algebra $\{Q,Q\}$. We thank Dr. Hai Lin 
for informing us of the paper \cite{Lin:2005nh}}
\begin{eqnarray}
 \delta X^I &=& i\bar{\epsilon} \Gamma^I \Psi, \nonumber \\
  \delta \Psi &=& (D_{\mu} X^I) \Gamma^{\mu} \Gamma_{I} \epsilon
   -\frac{1}{6} [X^I,X^J,X^K] \Gamma_{IJK} \epsilon - \mu
   \Gamma_{3456}\Gamma^{I}X^{I}\epsilon, \nonumber \\
 \delta \tilde{A}_{\mu \ a}^{\ b} &=& i\bar{\epsilon} \Gamma_{\mu}
  \Gamma_{I} X^I_c \Psi_d f^{cdb}{}_a.
\end{eqnarray}
This deformed theory breaks the original $SO(8)$ $R$-symmetry down to
$SO(4) \times SO(4)$. By setting $\mu \rightarrow 0$ both the action and SUSY transformation
reduce to the original Bagger-Lambert action. In addition there is
another supersymmetry transformation:
\begin{eqnarray}
 &&\delta X^I_a = 0, \hspace{2em} \delta \tilde{A}_{\mu \ a}^{\ b} = 0,
  \nonumber \\
 &&\delta \Psi = \exp\left(-\frac{\mu}{3}\Gamma_{3456}\Gamma_{\mu}
  x^{\mu}\right)T^{-1}\eta,
\end{eqnarray}
where $x^{\mu}$ is the coordinates of the world volume.
In the massless limit of $\mu \rightarrow 0$, this becomes a constant shift
of the fermion $\delta \Psi =T^{-1} \eta.$ These inhomogeneous
supersymmetries correspond to the spontaneously broken supersymmetries
in $d=11$ by the presence of M2 branes. As in the case of D-brane effective
theories, they will play an important role in the full $d=11$ superalgebras
with 32 supercharges. 
\subsection{Deformed BL to Janus}
This model can be similarly investigated by expanding
the fields into modes with internal indices $a=(-1,0,i).$
The mode expansions of the mass and the flux terms become
\begin{eqnarray}
 {\cal L}_{mass} = \mu^2 X_{-1}^I X_0^I - \frac{\mu^2}{2} \hbox{Tr}
  (\hat{X}^I,\hat{X}^I) -i \mu \bar{\Psi}_{-1} \Gamma_{3456} \Psi_0 +
  \frac{i}{2}\mu \hbox{Tr}(\bar{\hat{\Psi}}\Gamma_{3456},\hat{\Psi}),
\end{eqnarray}
and 
\begin{eqnarray}
 {\cal L}_{flux} = \frac{2 i}{3}\mu \epsilon_{EFGH} X_0^E \hbox{Tr}
  (\hat{X}^{F},[\hat{X}^G,\hat{X}^H]) + \frac{2 i}{3}\mu
  \epsilon_{E^{'}F^{'}G^{'}H^{'}} X^{E^{'}}_0 \hbox{Tr}(\hat{X}^{F^{'}},[\hat{X}^{G^{'}},\hat{X}^{H^{'}}]).
\end{eqnarray}
Now $X^{I}_{-1}$ and $\Psi_{-1}$ again appear linearly in the action, 
and they are Lagrange multipliers. 
Because of the mass terms, the constraint
equations are modified to 
 \begin{eqnarray}
  (\partial^2 - \mu^2)X^I_0 = 0, \hspace{2em} (\Gamma^{\mu}
   \partial_{\mu} + \mu \Gamma_{3456})\Psi_0 = 0.
 \end{eqnarray}
 Namely the fields with the $T^0$ component are constrained to 
 obey  the massive wave equations.
 Since $X^I$ are real fields, instead of the plane waves
 $\exp(i k_\mu x^\mu)$ with a time-like vector $k_\mu$, 
 we take the following solution to the constraint equation;
 \begin{eqnarray}
  X^{I}_0 = f e^{p_{\mu}x^{\mu}} \delta^{I}_{10} = v(x) \delta^{I}_{10},
   \hspace{1.5em} \Psi_0 = 0,
 \end{eqnarray}
 where $f$ is an arbitrary constant and $p_{\mu}$ is a spacelike vector
 satisfying $p^2 = \mu^2$. Without loss of generality, we
 can take $p_\mu=(0,\mu,0)$.
 This configuration  preserves half of the 16 supersymmetries, 
 since $\Psi_0$ transforms as:
 \begin{eqnarray}
  \delta \Psi_0 = v(x) \mu ( \Gamma^{1} - \Gamma_{3456})
   \Gamma^{10} \epsilon.
 \end{eqnarray}
Hence around the above configuration, we will get  Janus gauge
field theories with 8 supersymmetries.
(For general solutions, more supersymmetries are broken.) 

Inserting this configuration to the action, one can again
 integrate the redundant gauge field $B_{\mu}$.
Terms involving $B_{\mu}$ are given by:
 \begin{eqnarray}
  {\rm Tr}\left[-\frac{1}{2} (\hat{D}_{\mu} \hat{X}^{10} - v B_{\mu})^2 +
   \frac{1}{2} \epsilon^{\mu\nu\lambda} \hat{F}_{\mu\nu} B_{\lambda}
   - p^{\mu} v B_{\mu} \hat{X}^{10}\right].
 \end{eqnarray}
 Integrating $B_{\mu}$ gives
 \begin{eqnarray}
  && {\rm Tr}\left[\frac{1}{2v}\epsilon^{\mu\nu\lambda}
   \hat{F}_{\mu\nu} p_{\lambda} \hat{X}^{10} + \frac{1}{8v^2}
   (\epsilon^{\mu\nu\lambda}\hat{F}_{\mu\nu} -
   2v\hat{X}^{10}p^{\lambda})^2 \right]\nonumber \\
  &&=  - \frac{1}{4v^2} {\rm Tr} \hat{F}_{\mu\nu}^2 +
   \frac{\mu^2}{2} \hbox{Tr} (\hat{X}^{10},\hat{X}^{10}).
 \end{eqnarray}
Interestingly the second term is canceled by the 
mass term of $\hat{X}^{10}$
and  all the terms  involving $\hat{X}^{10}$ have
 disappeared.  
 To summarize, the resultant  effective Lagrangian is given by:
\begin{eqnarray}
{\cal L} &=& -\frac{1}{2} \hbox{Tr} (\hat{D}_{\mu} \hat{X}^{A})^2
   -\frac{\mu^2}{2} \hbox{Tr}(\hat{X}^A,\hat{X}^A) +
   \frac{1}{4} v^2 [\hat{X}^A,\hat{X}^B]^2 \nonumber \\
   & & + \frac{i}{2} {\rm Tr}\left(\bar{\hat{\Psi}}\Gamma^{\mu} \hat{D}_{\mu} \hat{\Psi}\right)
    + \frac{i}{2}\mu \hbox{Tr}(\bar{\hat{\Psi}}\Gamma_{3456},\hat{\Psi})
    +\frac{1}{2} v {\rm Tr}\left(\bar{\hat{\Psi}} 
    [\hat{X}^A,\Gamma_{10, A}\hat{\Psi}]\right)
    -\frac{1}{4v^2} {\rm Tr}\hat{F}^2_{\mu\nu} \nonumber \\
    & & - \frac{2 i}{3} v \mu \epsilon^{A^{'}B^{'}C^{'}10}
    \hbox{Tr}(\hat{X}^{A^{'}},[\hat{X}^{B^{'}},\hat{X}^{C^{'}}]).
\end{eqnarray}
This is a Janus field theory whose coupling constant is given
by  $v=f \ \exp(\mu x^1)$. 
The Lagrangian is invariant under the following 8 supersymmetries
\begin{eqnarray}
 \delta \hat{X}^A &=& i \bar{\epsilon}\Gamma^A \hat{\Psi},\notag\\
 \delta \hat{\Psi} &=& \hat{D}_\mu \hat{X}^A \Gamma^\mu \Gamma^A
  \epsilon
  - \frac{1}{2 v} \epsilon_{\mu\nu\lambda}\hat{F}^{\nu\lambda}\Gamma^\mu
  \Gamma^{10}\epsilon
  +\frac{i}{2} v[\hat{X}^A, \hat{X}^B]\Gamma^{AB}\Gamma^{10}\epsilon
  -\mu \Gamma_{3456} \Gamma^A \hat{X}^A \epsilon,\notag\\
 \delta \hat{A}_\mu &=& i v\bar{\epsilon}\Gamma_\mu \Gamma^{10} \hat{\Psi},
\end{eqnarray}

Finally if $v$ vanishes, i.e. for $X^{I}_0 = 0$ and $\Psi_0=0$,
the  Lagrangian becomes
\begin{eqnarray}
{\cal L} &=& -\frac{1}{2} {\rm Tr}(\hat{D}_{\mu} \hat{X}^I)^2
 + \frac{i}{2} {\rm Tr}\left(\bar{\hat{\Psi}} \Gamma^{\mu} \hat{D}_{\mu}
                        \hat{\Psi} \right)
 - \frac{\mu^2}{2} \hbox{Tr} (\hat{X}^I,\hat{X}^I) + \frac{i}{2}\mu
  \hbox{Tr}(\bar{\hat{\Psi}}\Gamma_{3456},\hat{\Psi}),
\end{eqnarray}
with a constraint $\hat{F}_{\mu \nu}=0$.
The supersymmetry transformation is given by 
\begin{eqnarray}
 \delta \hat{X}^I &=& i \bar{\epsilon} \Gamma^I \hat{\Psi}, \notag\\
 \delta \hat{\Psi} &=& \hat{D}_{\mu} \hat{X}^I \Gamma^{\mu} \Gamma^I
  \epsilon - \mu \Gamma_{3456} \Gamma^I \hat{X}^I \epsilon, \notag\\
 \delta \hat{A}_{\mu} &=& 0
\end{eqnarray}
and the Lagrangian has the  $SO(4) \times SO(4)$ R-symmetry.

\section{Conclusions and discussions}
\setcounter{equation}{0}
In this paper, we have derived Janus field theories from the Bagger-Lambert
field theory with the specific realization of 3-algebra given by
\cite{Gomis:2008uv,Benvenuti:2008bt,Ho:2008ei}. By integrating redundant fields,
we obtained supersymmetric field theories whose 
coupling  varies with the space-time coordinates.
A similar analysis was also done for the mass-deformed 
Bagger-Lambert model. 
In this case, we obtained a mass-deformed supersymmetric
Yang-Mills theory with an exponentially growing coupling
constant along one of the spacial direction.

The analysis in this paper became possible  by the 
remarkable discovery of the realization of the 3-algebra.
The roles played by the fields associated with the internal indices
$T^{-1}, T^0$ and $T^i$ are completely different, 
and this is the origin  of the success that the D2 brane effective 
theory can be reproduced from the very strangely looking model
of Bagger-Lambert.  

One of the most important directions 
will be to construct 
 a matrix model of M-theory with $SO(10,1)$ symmetry.
 In the case of matrix models for superstrings, a superstring world sheet action 
is related to the D-brane gauge theories through matrix models
\cite{BFSS,IKKT}.
Similarly we may expect that the supermembrane world volume
action must be related to the Bagger-Lambert gauge theories of multiple
M2-branes through a new class of matrix models.
 A natural guess  \cite{IUZ} is 
 \begin{equation}
 S =  {\rm Tr}\left( -\frac{1}{6} [X^I, X^J, X^K]^2 + \frac{1}{2} \bar{\Psi} \Gamma_{I, J} [X^I, X^J,\Psi]\right),
 \end{equation}
where $I$ runs from $0$ to $10$, but the action is not invariant
under  supersymmetry transformations. 
This action is closely related to both of the supermembrane action and 
the Bagger-Lambert action, but unfortunately it seems 
different from both of them.
The difficulty in the supermembrane action is that we cannot fix the 
$\kappa$-symmetry without breaking $SO(10,1)$ rotation. 
The difficulty to construct a gauge theory is how to exactly identify
the gauge fields  of the Bagger-Lambert model and its
supersymmetry transformation in terms of the matrix model. 
The recently discovered 3-algebraic structure suggests 
that the embedding of the space-time in the internal space
is more complicated than the case of the matrix models (i.e. large N reduction). 
We want to come back to this problem in near future.

\section*{Acknowledgments}
We thank Drs. Y. Hikida and H. Umetsu for discussions
and Dr. Umetsu for collaborations on a construction of 
matrix models for M-branes and M-theory.

\end{document}